\documentclass{vgtc}                          

\ifpdf
  \pdfoutput=1\relax                   
  \usepackage{graphicx}                
  \DeclareGraphicsExtensions{.pdf,.png,.jpg,.jpeg} 
\else
  \usepackage{graphicx}                
  \DeclareGraphicsExtensions{.eps}     
\fi%

\graphicspath{{figures/}{pictures/}{images/}{./}} 

\usepackage{microtype}                 
\PassOptionsToPackage{warn}{textcomp}  
\usepackage{textcomp}                  
\usepackage{mathptmx}                  
\usepackage{times}                     
\usepackage{cite}                      
\usepackage{tabu}                      
\usepackage{booktabs}                  

\usepackage{xcolor}
\usepackage{colortbl}
\usepackage{arydshln}
\definecolor{someColor}{RGB}{210,205,255}
\definecolor{greyColor}{RGB}{240,240,248}
\definecolor{nCol}{RGB}{253,193,204}
\definecolor{pCol}{RGB}{210,205,255}
\definecolor{mCol}{RGB}{249,249,255}
\onlineid{0}

\vgtccategory{Research}

\vgtcinsertpkg

\title{Micro-entries: Encouraging Deeper Evaluation of Mental Models Over Time for Interactive Data Systems}

\author{Jeremy E. Block\thanks{e-mail: j.block@ufl.edu} %
\and Eric D. Ragan\thanks{e-mail:eragan@ufl.edu}}
\affiliation{\scriptsize Department of Computer \& Information Science \& Engineering\\ University of Florida}

\teaser{
  \centering
  \includegraphics[width=0.95\linewidth]{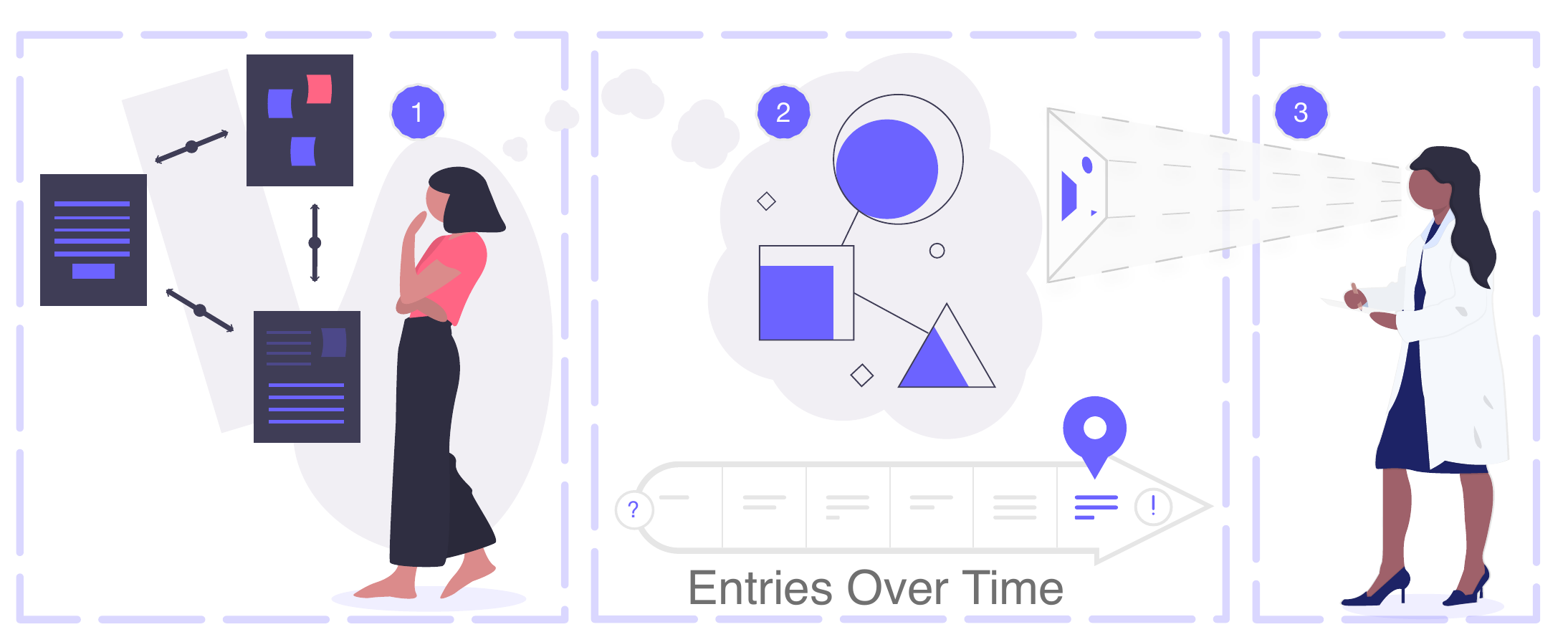}
  \caption{A graphical representation of capturing and evaluating a user's mental model during system interaction. 
  As users work with an application (1), they are asked to describe noticed patterns and provide their explanations many times (2), which can then be studied by researchers (3).
  This approach can encourage users to do more reflection of the system performance to reach more informed and less impressionable understandings of system limitations. 
  This capture method provides more comprehensive and higher fidelity accounts of the user's mental model while also tracking how it changes over time.
  }
  \label{fig:teaser}
}

\abstract{

Many interactive data systems combine visual representations of data with embedded algorithmic support for automation and data exploration.
To effectively support transparent and explainable data systems, it is important for researchers and designers to know how users understand the system.
We discuss the evaluation of users' mental models of system logic.
Mental models are challenging to capture and analyze.
While common evaluation methods aim to approximate the user's final mental model after a period of system usage, user understanding continuously evolves as users interact with a system over time.
In this paper, we review many common mental model measurement techniques, discuss tradeoffs, and recommend methods for deeper, more meaningful evaluation of mental models when using interactive data analysis and visualization systems.
We present guidelines for evaluating mental models over time that reveal the evolution of specific model updates and how they may map to the particular use of interface features and data queries.
By asking users to describe what they know and how they know it, researchers can collect structured, time-ordered insight into a user's conceptualization process while also helping guide users to their own discoveries.

} 

\CCScatlist{
  \CCScatTwelve{Human-centered computing}{Visu\-al\-iza\-tion}{Visualization design and evaluation methods}{}
}

\begin{document}


\maketitle

\section{Introduction} 

Interactive data systems permeate numerous contexts and facets of life with advances in algorithm-supported tools to assist humans in data analysis and decision making.
Artificial intelligence and machine learning algorithms are becoming ubiquitous due to their versatile pattern matching abilities and general superior performance at highly specific tasks.
The visual analytics community continuously innovates new technology and solutions to data problems with automation to help simplify complex data, reveal patterns of interest, or recommend potentially relevant information.
Yet, concerns remain surrounding how model reliability, uncertainty, and bias might affect the quality and validity of decision making~\cite{miller2019explanation,Doshi-Velez_Kim_2017}.

Practical applications of algorithms require system designs that help users understand the underlying logic of automated support.
In many contexts, end users avoid taking advantage of the algorithmic support that they do not understand~\cite{dietvorst2016people,dasgupta2016familiarity,goodall2018situ,Hosanagar_Jair_2018};
in other cases, users may over-rely on automation without a critical mindfulness of the system's limitations and flaws~\cite{Nourani_Honeycutt_Block_Roy_Rahman_Ragan_Gogate_2020,cummings2004automation,parasuraman1997humans}.
There is a growing need for researchers and designers to understand how users perceive system functionality.
Aiming to achieve \textit{transparency} and \textit{interpretability}, \textit{explainable AI} (XAI) research has turned to visualization techniques as a potential antidote for elucidating the metaphorical black-box that is machine learning~\cite{Mohseni_Zarei_Ragan_2018,arrieta_2020,Herman_2019,Hohman_Park_Robinson_Chau_2019}. 
However, without robust ways to measure what is comprehended about the models and algorithms, claims of achieving interpretability are limited~\cite{Lipton_2017,Doshi-Velez_Kim_2017}.

Moreover, meaningful evaluation of human understanding is challenging~\cite{miller2017explainable, miller2019explanation}.
Human evaluation is essential to assess interpretability and produce actionable design knowledge, but this requires researchers to find ways to peer inside the heads of those who use data systems~\cite{Hoffman_Mueller_Klein_Litman_2019,Klein_Militello_2001}.
For most studies that evaluate human understanding, researchers often rely on numerical self-reported measures of trust administered once at the end of the session~\cite{Schmidt_Biessmann_2019}
or throughout the task~\cite{honeycutt2020solicit,nourani2020Role}.
While easy to administer, such simple measures lack the ability to accurately assess how specific system features and elements of visual design influence user understanding.

In this paper, we discuss the benefits and tradeoffs of more comprehensive evaluation methodology for assessing users'  \textit{mental models}~\cite{Craik_1943} of how intelligent data systems function.
While much research in visualization and human-computer interaction has studied the state and evolution of users' thinking and sensemaking during data analysis, we provide particular attention to the study of users' mental models of applications' algorithmic capabilities.
To this end, we provide grounding for a methodology that encourages more thoughtful participation and insightful data capture.
Drawing on existing research that provides a variety of methods, our goal is to rationalize a need for more meaningful evaluation of mental models for interactive analysis systems.
In particular, we discuss the importance of tracking the progression of user thinking over time with attention to cause-and-effect relationships between specific system features and user-defined understanding.
By recognizing how specific user interactions with system design elements contribute to updates in the user's mental model, researchers can better design supportive, understandable and intelligent systems.
We present implementation recommendations for the discussed methodology along with a use case in an explainable visualization application.

\section{Evaluating User's Mental Models}\label{background}
For our discussion of evaluating user understanding of interactive data systems, we follow prior work in using the term \textit{mental model} to refer to a personal representation of the world, its relationships, and their subsequent interactions.
Expanding on Craik's theory of miniature worlds~\cite{Craik_1943}, many more nuanced definitions of mental models and their construction have been proposed~\cite{vanderVeer_Puerta_Melguizo_2002}. 
Some claim that mental models take on different forms and levels of fidelity; from a surrogate of the world to a complex network of the system state and possible actions available to users~\cite{Carroll_Olson_1987}.
Rouse and Morris theorize a similar functional understanding: stating that mental models describe, explain, and predict a system's purpose, form, function, and state~\cite{Rouse_Morris_1985}.
Norman considered mental models to be unstable over time, emotionally and superstitiously charged, and often only limited to the smallest subset of all things communicated by the interface\cite{Norman_1983}.
More contemporary beliefs suggest that the construction of a mental model and the reasoning with that mental model are two disjoint systems~\cite{Johnson-Laird_2010}. 
While the mental model reduces the load on working memory, the process of confirming our understandings and arriving at a conclusion requires reasoning.
Reasoning helps us find the holes in our understanding, as evidenced by the Illusion of Explanatory Depth: \textit{``[A] theory that seems crystal clear and complete in our head suddenly develops gaping holes and inconsistencies when we try to set it down on paper''}~\cite{Rozenblit_Keil_2002}.
It is clear that mental model construction is intuitive for people but asking users to consciously reason through \textit{how} their model works is less common.




The realm of education commonly employs multiple assessments to evaluate learning and mental model improvement over time~\cite{Isaacson_Fujita_2006,Bull_Ginon_Boscolo_Johnson_2016,Baldwin_2000}.
Drawing from their methodology, the use of \textit{reflective journaling} techniques can benefit learners. 
Students can track progress, reinforce key concepts, and review notes to anchor their understanding. 
Thus, asking users to write out and refine their mental models could provide researchers more succinct representations of what is learned from an interface while also elucidating the `aha' moments over time.

Many visualization researchers recognize the difficulty in conducting evaluations that capture a participant's comprehension~\cite{Lam_Bertini_Isenberg_Plaisant_Carpendale_2011,Ellis_Dix_2006,North_Chang_Endert_Dou_2011}.
Lam et al.~\cite{Lam_Bertini_Isenberg_Plaisant_Carpendale_2011} describe seven unique scenarios with different evaluation questions designed to target various goals.
When researchers want to understand what is communicated through a visualization, they recommend evaluating the interface in a controlled experiment with post-task learning assessments or interviews.
These techniques are beneficial because they are relatively unobtrusive to the participant during the evaluation and conclude with the participant explaining their behaviors, impressions and understanding~\cite{Lam_Bertini_Isenberg_Plaisant_Carpendale_2011}. 
Yet, participants have diverse communication abilities and imperfect memories, so relying on accurate post-task retelling as the only measure is risky.
Drawing from ethnographic and sociological research, Carpendale~\cite{Carpendale_2008} encourages greater variety in visual interface evaluations.
By incorporating multiple methods, the validity of visualizations can be confirmed while also enabling deeper exploration into the underlying phenomena of results~\cite{Carpendale_2008}.
North et al.~\cite{North_Chang_Endert_Dou_2011} discuss the importance of studying how users arrive at their conclusions following \textit{analytic provenance}, which can uncover interesting discussions into potential behavioral differences between user groups~\cite{Saraiya_North_VyLam_Duca_2006}.
The technique described in this paper adopts similar priorities, but adds particular attention to capturing the development of mental models for the underlying algorithms and analytic models behind data systems.
Since mental models are fluid and evolve with experience, the micro-entries approach emphasises tracking how a user's understanding of the system model updates over time.

Before outlining practical methodological recommendations for interactive data applications, this section first provides an overview of common methods used to capture mental models\footnote{The discussed collection of methods is non-exhaustive; refer to Hoffman et al. for further review~\cite{Hoffman_Mueller_Klein_Litman_2019}.}.
In particular, we focus on their ability to significantly capture temporal change and capture user reasoning for understanding of algorithmic capabilities in analysis applications with integrated machine learning.


\subsection{Quantitative Methods}
    Quantitative methods have the benefit of supporting simplified interpretations and comparable results.
\subsubsection{Matching Mental Models}
    Because mental models tend to be fuzzy~\cite{Norman_1983}, providing some clear examples helps users bring clarity to their interpretation.
    Typically, a constrained task is presented to users that helps to dissect the cognitive processes underlying their mental state~\cite{Klein_Militello_2001}.
    The \textit{matching mental models} method asks users to select an explanation or diagram that is the ``Nearest Neighbor'' to their beliefs.
    From these selections, the researcher can estimate understanding via a discrete and quantifiable measure.
    For example, Hardimam et al.~\cite{Hardiman_Dufresne_Mestre_1989} gave physics students a problem and asked them to select an alternative problem that would be solved most similarly. 
    With careful attention to the alternatives provided, the researchers could determine if participants decided similarity based on simply the surface features of the problem or a deeper understanding of the analogous physics formulas employed in solving it.
    By comparing user responses to a reference model, this method allows for a more operational assessment of mental model
    and has been applied to other contexts including educational games~\cite{Wasserman_Koban_2019}, and graduate coursework~\cite{GlazerWaldman_Cox_1980}.
    Typically assessed once at the end of a session, Glazer-Waldman and Cox\cite{GlazerWaldman_Cox_1980} show how it has been adapted to assess students throughout their course.
    
\subsubsection{Prediction Tasks}

    For practical uses of algorithms and intelligent systems, users want to understand how systems work in order to know when they can trust and rely on their results.
    Thus, considering the user's ability to correctly predict system output for a given input provides a meaningful and concrete measure of user understanding.
    If users can correctly (and consistently) predict system outputs, it follows that their mental model should be more complete; whereas, discrepancies between expectations and actual system function indicate limited user understanding.
    Many studies have employed variations of prediction tasks for assessing user understanding of systems (e.g.,\cite{nourani2019effects,Nourani_Honeycutt_Block_Roy_Rahman_Ragan_Gogate_2020,Schaffer_ODonovan_Michaelis_Raglin_Hollerer_2019,Poursabzi-Sangdeh_Goldstein_Hofman_Vaughan_Wallach_2019}).
    While practical, simple prediction tasks might not account for assessing if users accurately understand \textit{why} the predicted result will happen.
    Use of prediction in combination with confidence ratings and free response elaborations can help address this limitation~\cite{Hoffman_Mueller_Klein_Litman_2019,Nourani_Honeycutt_Block_Roy_Rahman_Ragan_Gogate_2020}.
    Typically, users are asked to predict system responses immediately after working with the system, but there is potential to ask users to predict system responses throughout the session and ask them to explain their responses too.

\subsection{Qualitative Methods}

While quantitative approaches allow for convenient and clean interpretation, they also tend to be highly specific and potentially overly simplified.
Actual human thinking and mental model development often has a messy nature, so qualitative methods allow greater flexibility in evaluating with higher levels of fidelity---but often at the cost of increased effort or complexity of data capture and interpretation.
Many methods draw on \textit{Thematic Analysis} to derive conclusions~\cite{Braun_Clarke_Hayfield_Terry_2018}.
In the following sections, we discuss some relevant methods and how they capture temporal data or encourage users to practice reflection.

\subsubsection{Think Aloud Methods}

    In Ericsson and Simon's original description, the think-aloud method introduced users to the process of vocalizing thoughts with an example, before working with the item of study uninterrupted~\cite{Ericsson_Simon_1984,Russo_Johnson_Stephens_1989}.
    They recommended only prompting users with simple reminders after a set waiting period expired~\cite{Ericsson_Simon_1984}.
    Individual comments were transcribed with their timings before being coded to reveal a proxy for individuals conceptualizations.
    In more recent iterations, methodological relaxations have been made, and Boren and Ramey argue this should only be done with justification~\cite{Boren_Ramey_2000}.
    One concern is that requiring or requesting updates can disrupt user thinking, which may reduce fidelity of the evaluation or interrupt the development of their mental model.
    There is also evidence that users stop talking when the cognitive load is high~\cite{Denning_Hoiem_Simpson_Sullivan_1990}; this should be an expected phenomena when users work with complex visualizations. 
    There is a balance between staying silent for users to describe their thoughts organically and prompting users to explain what they know.
    

\subsubsection{Interviews}
    Interviews with users, after interacting with systems, is a common practice to elicit overall user perspective and general understanding.
    Interview questions can be structured or unstructured, while also targeting specific topics or asking users to reflect on the overall experience~\cite{Merton_Kendall_1946}.
    Alternative versions of this technique ask users to ``teach back'' what they understand to an imaginary person.
    The accuracy of their statement helps to communicate their personal understanding~\cite{Hadinh_Bonner_Clark_Ramsbotham_Hines_2016}.
    Typically interviews are agnostic to temporal phenomena, and this lack of context can lead users to overgeneralize and potentially forget key details.
    In addition to post-task interviews, Klein and Mitello describe \textit{cognitive task analysis}, which attempts
    to derive the cognitive skills involved with a task~\cite{Klein_Militello_2001}. 
    Generally, experts are prompted to repeatedly walk through the decision making steps in higher and higher detail to help researchers identify strategies these experts have learned to use in their domain. 
    This is typically a challenging interview to conduct but can lead to valuable insights into the cognitive processes of experts.

\subsubsection{Retrospective Walk-through}
    Similar to the interview method described previously, retrospective walk-throughs invite users to watch a replay of their activities and provide explanations for their behaviors~\cite{Kuusela_Paul_2000}.
    Extending this method with an alternative question (``when did you make the mistake?'') is known as fault diagnosis, and helps to identify faulty reasoning and misconceptions~\cite{Puerta-Melguizo_Chisalita_VanderVeer_2002}. 
    As participants review their performance, temporal signifiers such as, ``at first I thought...'' or ``before I recognized...'', can give clues to how models develop, but are not liberated from the fabrication and forgetfulness of events~\cite{Russo_Johnson_Stephens_1989}.
    Research has demonstrated that participants can quickly forget or incorrectly remember details about their thinking and process when interacting with data systems~\cite{ragan2015evaluating}, though the approach can be especially useful for clarifying observed events or ambiguous results captured through other methods~\cite{Ragan_Goodall_2014}.



\subsubsection{Diary Studies}
    Traditionally used to help find patterns in longitudinal use cases, the diary study asks users to record thoughts and impressions as they experience them~\cite{Hart-Davidson_Spinuzzi_Zachry_2007}.
    Common uses are aimed at capturing the frequency of events~\cite{Vrotsou_Ellegard_Cooper_2007} or encouraging reflection on phenomena with prompts to explain and improve behavior pattern recognition~\cite{Walker_2006}.
    When prompted, participants can be more considerate of how they see the world and---importantly---describe their perspective in their own words.

    
    
\subsubsection{Concept Mapping}
    Many forms of illustrative system and diagramming techniques have been described for approximating and expressing mental models~\cite{Nakatsu_2009_3,Evans_Jentsch_Hitt_Bowers_Salas_2001}.
    A common form is to use concept maps as either formal or informal graph representations consisting of boxes and lines. 
    Cabrera et al.~\cite{Cabrera_Cabera_Sokolow_Mearis_2018} claim that visual mapping empowers thinkers to symbolize their ideas, interconnect their parts, and manipulate them tangibly---referring to this as a tool of the systems thinker.
    In a later work, they define the goal of systems thinking as increasing ``the probability that our mental models are in alignment with reality''~\cite{Cabrera_Cabrera_Troxell_2020}.
    Their process utilizes simple rules for mapping complex mental models and suggest that this flavor of system's modeling will inherently support the deconstruction of phenomena. 
    Unfortunately, their technique---while simple in foundation and indented to mimic the natural cognitive process---requires practice and instruction to implement and utilize effectively.
    While some may find these visual techniques to be more in line with their personal mental modality, written words can be generic, familiar, offer flexibility for structure and are space efficient.



\begin{table}[t]
\centering
\begin{tabular}{|r|c|c|}
\hline
\textbf{Method} & \textbf{Temporal} & \textbf{Reflection} \\ \hline
Matching & \cellcolor{mCol}o & \cellcolor{nCol}- \\ \hline
Prediction & \cellcolor{mCol}o & \cellcolor{mCol}o  \\ \hline
Think Aloud & \cellcolor{pCol}+ & \cellcolor{mCol}o \\ \hline
Interviews & \cellcolor{nCol}- & \cellcolor{pCol}+ \\ \hline
Walk-through & \cellcolor{pCol}+ & \cellcolor{pCol}+ \\ \hline
Diary Studies & \cellcolor{pCol}+ & \cellcolor{pCol}+ \\ \hline
Concept Maps & \cellcolor{mCol}o & \cellcolor{pCol}+ \\ \hline
Microgenetic & \cellcolor{pCol}+ & \cellcolor{pCol}+ \\ \hline
\end{tabular}
\vspace{0.2cm}
\caption{A glanceable summation of mental model measurement methods. Temporal features record time as a typical component of the method, whereas reflection generally inspires users to reason through their understanding. Here (+) denotes the feature's presence, whereas (-) represents the lack of feature. Methods that could adopt the feature are marked with a (o).}
\label{tab:mm_sum}
\end{table}






\subsubsection{Microgenetic studies}
The microgenetic method is traditionally applied to children's cognitive development or problem solving~\cite{Flynn_Pine_Lewis_2006}, but has also seen promise in graduate level medical education~\cite{Smith_Corrigan_2018}.
The technique assumes that 1) some knowledge will be gained during the observation, 2) the researchers assess more often then the knowledge is gained, and 3) the observed behavior is not impacted by the measurement technique.
When a key insight takes place---because of the frequent evaluation---the potential reasons why the change occurred can be uncovered based on the conditions of the situation before and after the change~\cite{Siegler_Crowley_1991}.
The assessment, while dependent on the context of the study, often consists of some rubric of expected learning outcomes (e.g., do they do basic multiplication by writing or in their head, etc.). 
Repeated reevaluations can be performed to estimate knowledge acquisition rate.
 On the other hand, because they require repeated assessments, concerns related to practice or boredom are often considered in parallel with the microgenetic technique.
 
\section{Micro-entry Evaluation of Mental Model Evolution in Interactive Systems}

Of the previously described mental model evaluation techniques typically used in research with interactive data systems, few ask users to consciously reason through their understanding while also capturing changes in understanding over time.
For example, presenting a concept map of how the system works may require thoughtful reflection of system performance, but typically only the final product is evaluated by researchers.
Alternatively, the think aloud method may capture changes in thought over time, yet traditionally the user is uninterrupted and free to explore without encouraged reflection on previous observations.
Research can benefit from greater adoption (and adaption) of methodology that encourages users to reflect on what they know and how it changes over time. 
We encourage a mental model evaluation method 
that captures patterns recognized by users and their rationale repeatedly, while also prompting users to reflect on previously held beliefs.

We refer to this technique as capturing through \textit{micro-entries}, a neologism derived from microgenetic research that emphasises repeated evaluation~\cite{Luwel_2012,Cheshire_Muldoon_Francis_Lewis_Ball_2007} and standardized, reflective diary entries.

\subsection{Theoretical Basis for Micro-entry Evaluation}

In education, reflection has been known to encourage novel and more meaningful insight for students~\cite{Hampton_Morrow_2003,Chi_2013,Isaacson_Fujita_2006,Boud_Keogh_Walker_Keogh_Walker_1985,Garcia_Chu_Nam_Banigan_2018,Siegler_2002}.
Yet, there's evidence that people are not aware of how to reason through systems and model ideas by default~\cite{Hung_Blumschein_2009}.
Asking users to conceptualize what they understand more deeply can override default behavior and lead to a more clear understanding; which, theoretically, facilitates the communication of their ultimate mental model. 
Similar to a student in a classroom, this belief can be extended to say that deep reasoning about what one knows is not normal/expected cognitive behavior in individuals presented with a new experiences but instead encouraged by reflection prompts such as journal entries, providing justifications, or answering a `why' question.
In fact, a prominent theory of mind is that there are two cognitive systems at play: one that hastily constructs mental models with intuitive explanations and another that methodically deliberates on the handful of concepts in working memory~\cite{Johnson-Laird_2010}.
By asking users to reason through what they know, we elicit responses from that latter, more methodical system.

We can expect that thoroughly reasoned mental models tend to be more grounded and consistent for users, leading to more confident responses when asked to predict system output, identify system weaknesses, or describe what they can trust the system to do.
At its heart, asking users to reason through their mental model may involve a form of sensemaking~\cite{Pirolli_Card_2005} through self reflection.
The approach is commonly included in cognitive task analysis ~\cite{Klein_Militello_2001} and elements of model-facilitated learning ~\cite{Milrad_Spector_Davidsen_2002}.
All three encourage users to reason through observations and describe how they construct knowledge architecture that explains observations.
When users are asked to consider what they know and how they know it, people naturally construct their own explanation from their past experience~\cite{Khemlani_Johnson-Laird_2011}.
This leads to conclusive and corrective understanding.
Micro-entries prompt individuals to explain their understanding of patterns (with a high sampling rate), encouraging more reflection on what they know, and how they know it.
By collecting what users notice in a list and asking for their explanation of the pattern over time, we encourage them to reevaluate and strengthen their understanding of the system while also communicating rich qualitative data tied to specific times and system phenomena.
With limited cognitive resources, we feel that user minds will benefit from available space to record what they are seeing and confront their beliefs of the system, which will, in turn, help researchers to more accurately understand the user's mental model. 
We propose incorporating flexible prompts to lead user discovery and serve as a tether when exploring open, complicated, and ill-defined ``interpretable'' spaces, while also providing researchers insight into what users believe at specific times.

\subsection{Implementing Micro-entries}

To guide practical use of the micro-entry approach, we propose the use of prompts to encourage users to reflect on their observations and summarize their understanding.
Repeatedly asking users, ``what pattern do you notice?'' followed with, ``how would you explain that pattern?'' will provide structured reasoning to their task~\cite{Klein_Militello_2001}. 
From most necessary, to least, we believe micro-entries should account for the following:

\begin{enumerate}

\item \textbf{Open data}: A way to record user ideas, which could allow for various representations or types of data collection (e.g., verbal, textual, spatial). 
A basic, yet consistent, method for recording qualitative notes and ideas is an obvious yet essential element.

\item \textbf{Frequent and time-stamped data}: The structured prompts should include time-relevant data to help reveal a participant's patterns over time.
Also, consider capturing the system state when entries are created or modified.

\item \textbf{Visible and accessible}: Users can manage past patterns by selecting them from a list and making edits.
These edits must be recorded and attributed to the original entry to extract a hierarchy of pattern shifts over time.

\item 
\textbf{Prompted reflection}: Participants should be prompted to explain their identified patterns or reevaluate previous explanations to reinforce what they can confirm or discredit anything incorrect.
Frequent reflection can lead to higher fidelity mental models while also showing development over time.

\item \textbf{Light-weight and unobtrusive}: Participants should be enabled to note a pattern, provide an explanation, and test that explanation quickly---ideally without additional barriers that may distract their attention.
A more responsive system, ensures more articulate and focused data capture.

\item \textbf{General or targeted capture}: 
Depending on the research and design goals, prompts and instructions can either prioritize specific types of understanding or allow more freedom and exploration. 
Telling users that they are `seeking to understand the system and communicate that understanding to the researchers' reinforces the use of micro-entries and helps users feel more informed.
We recommend explicitly telling users of their role to help establish an appropriate mindset: aiming to achieve a clear understanding of the system's performance.
\end{enumerate}

Of course, there are also a few concessions to examine when implementing micro-entries:
\begin{enumerate}
\item \textbf{Demanding feedback}: It is important to consider that the act of requesting explanations for the system's behavior establishes a \textit{demand structure} where participants may, ``feel compelled to give a reason, even if they did not have one prior to your question''\cite{Norman_1983}. 
Users may also feel expected to provide evidence of a mental model that matches your expectations and not representative of their own beliefs.
\item \textbf{Interrupting prompts}: Asking a user to review a previous theory or describe a new pattern while they are testing the fidelity of another observation can disrupt the user's cognitive process---and future discoveries may suffer.
Consider ways to control when users receive a prompt (visible timer, after specific interactions, at predefined moments in the task, etc.) or choose to prompt in a more subtle way (collapsible list, specific keystroke, raising a hand) to maintain free exploration.
\item 
\textbf{Offloading Working Memory}: Simply providing a space to describe noticed patterns may facilitate mental model evolution in different ways from studies that do not provide this feature during interaction.
The progressive evolution of a clearer and more solid mental model is the intention of the described approach, but its effects have yet to be compared to alternative mental model evaluation techniques.
Thus, we do not know if the user discoveries are a result of reflection or the offloading of working memory into an interface element. 
Future research will need to compare insights generated via micro-entries and other techniques.
\end{enumerate}

The micro-entry method lends itself especially well to written records stored at the periphery of the screen but the fundamental elements (discussed above) simply encourage users to reason with their mental models.
Our use case focuses on a diary-like method to demonstrate its utility.


\subsection{Interpreting Micro-entry Results}\label{interpretations}

Because micro-entries capture both semantic understanding and temporal relations, a number of data analysis approaches are appropriate candidates to assist in their interpretation.
With the goal of extracting not only one's final understanding, but also changes in identified patterns over time, these time-specific rationale can be revealed. 

The analysis of micro-entries may provide relevant insight to the interpretability questions proposed by Doshi-Velez and Kim~\cite{Doshi-Velez_Kim_2017}.
Here, we present simplified variations; appending an additional question considering the temporal dimension captured by micro-entries:
\begin{enumerate}
\itemsep-0.2em 
    \item What form do the noticed patterns take?
    \item How many patterns were noticed?
    \item How are patterns structured or aggregated?
    \item Is there evidence that users understand how patterns are related?
    \item How well do users understand the uncertainty in the system's responses?
    \item How do the patterns change over time? 
\end{enumerate}




Guided by such questions, analysis of micro-entry data will typically require qualitative analysis to extract a user's mental model and its changes.
In this section, we illustrate the potential ways micro-entries may offer insights, but these techniques are not prescriptive, as selection of best methodology will depend on the project specifics and implementation details when considering the pros and cons of any given approach.
\textit{Thematic Analysis}, a common approach for qualitative analysis, can be especially useful when tracking patterns and changes in users' mental models. 
\textit{Reflexive} thematic analysis recognizes that the researcher plays a part in the conceptualization of themes and can provide rich interpretations grounded in the collected data from iterations of review~\cite{Braun_Clarke_2006}.
Typically, there are six non-linear, recursive phases of reflexive thematic analysis: 1) familiarization, 2) code generation, 3) thematic prototyping, 4) prototype revising, 5) theme defining, and 6) report producing~\cite{Braun_Clarke_Hayfield_Terry_2018}.
Familiarization typically begins the thematic analysis and requires an inductive, open-minded reading of the data; progressing through rounds of inductive and deductive coding.
One benefit of micro-entries is that users repeatably describe what they notice and explain their rationale.
This makes latent codes easier to distill because the participant provides foundation to their claim.

On the other hand, due to their structured nature, defining a codebook and counting the frequency of specific events or identified patterns is also possible~\cite{Kulesza_Stumpf_Burnett_Wong_Riche_Moore_Oberst_Shinsel_McIntosh_2010}.
Drawing from microgenetic techniques, there will likely be a series of common and uncommon observations made by users.
Agreeing on a set of themes, defining a rubric and grading the users at each time step can reveal when key insights were discovered and also the relative rates of discovery. 
Alternatively, if the micro-entry method is used in a more exploratory analysis, the order of what participants choose to write about may uncover their flow of attention. 
Furthermore, extracting and visualizing the sentiment of topics (e.g.,~\cite{Wang_Xiao_Liu_Xu_Zhou_Zhang_2013}) may also be relevant when considering user understanding. 
By considering sentiment analysis of user reflections, user perceptions of the system capabilities, as well as their comforts and frustrations can be discovered.
Ultimately, micro-entries can be counted and analysed to reveal the proportion of incorrect or correct observations about system functionality and extract perceptual rates over time.

Since having semi-structured, qualitative, time-series, user-defined (and refined) interpretations can be overwhelming and complex, the next section discusses how visualization could be considered to help extract key insights.

\begin{figure*}[ht]
\centering
  \setlength{\belowcaptionskip}{-15pt}
  \includegraphics[width=1
  \linewidth]{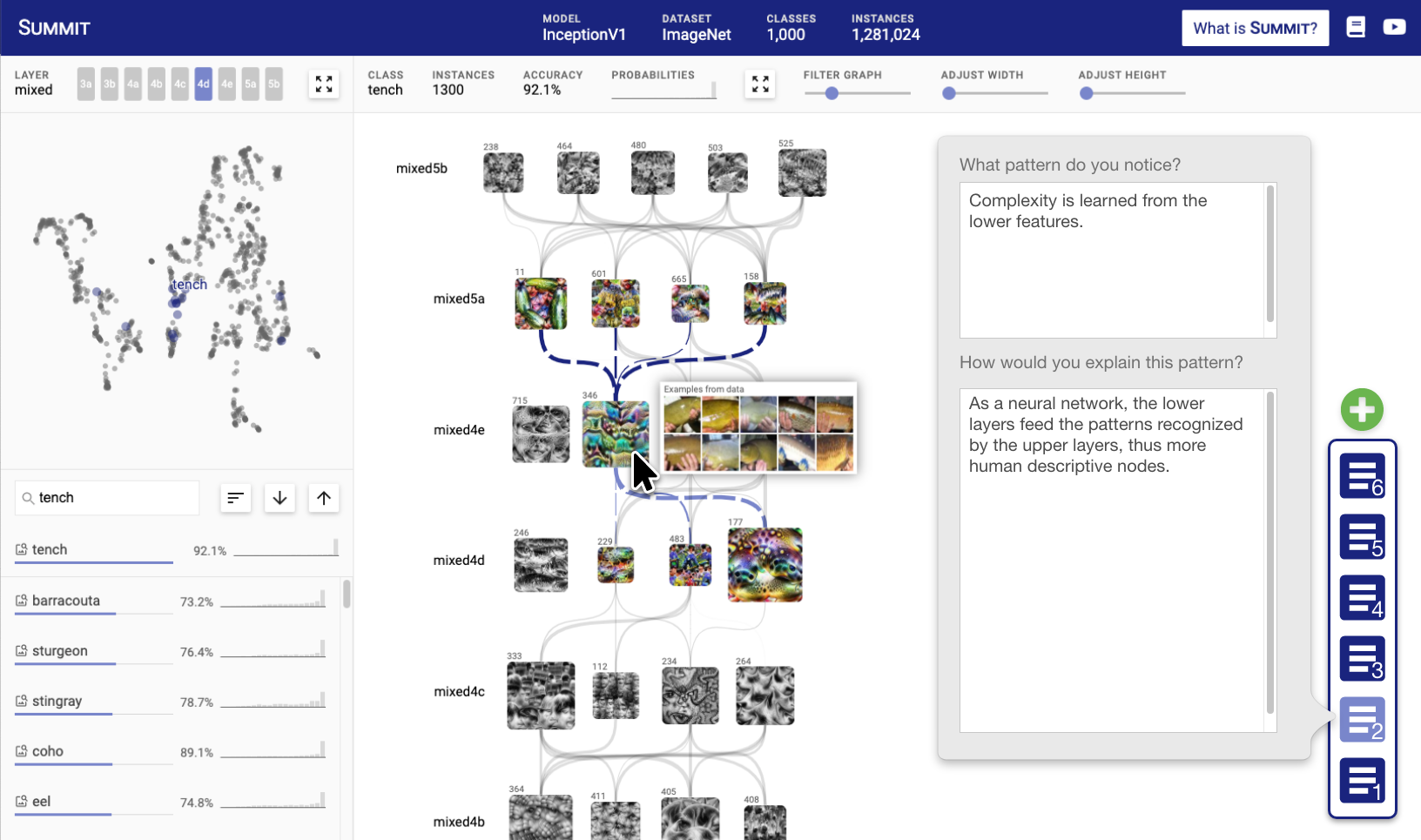}
  \caption{
  The proposed design concept applied to an existing explainable AI tool called `Summit' by Hohman et al.~\cite{Hohman_Park_Robinson_Chau_2019}.
  The micro-entry approach adds a journaling capability to the periphery of the main visualization.
  The user's identified patterns and justifications are recorded and collapsed into the icons at the right.
  Users can refine the entries as they work, or the tool will prompt them to reconsider random entries periodically.}
  ~\label{fig:tool}
\end{figure*}

\subsection{Visualizing Micro-entry Data}

Visualization can assist in analysis and interpretation of evolving mental models, and prior research has contributed many viable metaphors for visualizing textual data~\cite{Gan_Zhu_Li_Liang_Cao_Zhou_2014}, data over time~\cite{aigner2007visualizing,Aigner_Rind_Hoffmann_2012}, and studying the progression of analytic provenance~\cite{ragan2015characterizing,herschel2017survey}.
As suggested by Ragan and Goodall, for further clarification of changes over time, researchers can combine user process representations with user review and further commenting for additional insights on user understanding~\cite{Ragan_Goodall_2014}.

Looking to explore qualitative data extracted from the timing of events, Slone describes the use of \textit{spectrum}~\cite{Slone_2009}. 
After creating a codebook, the technique could be adapted to represent themes as rings with participants arranged around the outside of a circle.
At the intersection of each participant's segment and thematic ring, some metric (the relative number of entries made, the correctness of the identified pattern, etc.) could be encoded with color.
By shifting the order of the participants according to some condition, patterns may emerge such as: ``Analysts make correct entries more frequently then others.''

But beyond typical frequency visualizations, like word clouds~\cite{Heimerl_Lohmann_Lange_Ertl_2014}, the data from micro-entries lends itself to alternative exploratory visualizations because it captures the branching nature of ideas over time.
Looking at data over time more generally, a visual design such as the \textit{ThemeRiver} approach ~\cite{Havre_Hetzler_Nowell_2000} can focus more on communicating themes and attention of users over time.
By capturing the frequency of ideas mentioned across all entries over time, the emergent pattern shows what individuals focus their attention on by what they spend the most time writing about. 
History flow visualizations can show changes to documents over time by encoding the size of the change vertically and its timing horizontally~\cite{Viegas_Wattenberg_Dave_2004}.
One benefit of our proposed method, is how it captures change in mental model.
Each time a micro-entry is edited, the change can be captured and attributed to it's original entry---leading to a hierarchy of changes.
Using existing visualization techniques for understanding changes in hierarchical themes over time, \textit{SplitStreams}~\cite{Bolte_Nourani_Ragan_Bruckner_2020} could be adapted to show these changes as shifts in mental model. 
For example, with minor manual merging of semantically similar entries (and perhaps some coding), a mental model could be represented as a stream and draw out horizontally with its margins representing changes in ideas.
By coding the themes described in an entry, vertical shifting between streams could show how ideas develop, shift, and converge over time.

An approach used in microgenetic research shows relative change over time.
By scoring each participant against some rubric at each assessment interval, researchers can quantify the relative clarity of one's mental model.
This instrument would be domain dependent, but as an example, the rubric may help quantify the accuracy of an identified pattern, the uniqueness of a conclusion, or its relative overlap with the amount of seen information.
Over time, these assessment scores expectantly improve and by visualizing these changes in a line-chart, relative rates can be compared between users~\cite{Cheshire_Muldoon_Francis_Lewis_Ball_2007}.
To validate a between subject rate of change, typically the assessments (or in this case, micro-entries) need to occur along the same time steps for all participants.
Controlling for an entry's request interval is a logical adjustment to the proposed used case.
Alternatively, graphical chain modeling~\cite{Edwards_2000} has been used to reveal the relativistic relationship of variables in complex problems~\cite{Foraita_Klasen_Pigeot_2008}.
In the context of micro-entries, multiple variables can be captured and compared to see how different elements may influence each other.
Applicably, during an exploratory visual analysis task, 
Reda et al.~\cite{Reda_Johnson_Papka_Leigh_2016} discuss the use of Markov chain models to abstractly 
predict the likelihood of a user's next exploratory action based on their current state.
Relying on similar data as provided by micro-entries, they code interaction logs and participant verbalizations into the key processes relevant to exploratory tasks (e.g., navigating, structural analysis, etc.).
The researchers then create a sequence of states for each user.
Based on the frequency of those state transitions, a probability distribution can be predicted by Markov chain models---where frequent transitions between states represent the core behaviors supported by the visualization.

We emphasize that the discussed options are only examples for consideration, as many different promising and applicable visualization techniques can be beneficial for understanding mental model states and changes over time.
Specific choices of representation and analysis will depend on the chosen method of data capture, the underlying system design, the model being studied, and the particular research goals.

\begin{table*}[!t]
\centering
\resizebox{\linewidth}{!}{
\def\arraystretch{1.3}

\begin{tabular}{|lllp{4.5cm}|p{10cm}|}

\multicolumn{1}{l}{\textbf{Time}} & \multicolumn{1}{l}{\textbf{ID}} & \multicolumn{1}{l}{\textbf{Ver}} & \multicolumn{1}{l}{\textbf{What do you notice?}} & \multicolumn{1}{l}{\textbf{How do you explain the pattern?}} \\ \hline \rowcolor{someColor}
\multicolumn{1}{|l}{0:00} &  & & \multicolumn{2}{l|}{Activity: Inspecting the class ``White Wolf''} \\ \hdashline
0:04 & 01 & a & ``Data flows from bottom to top'' & ``The animation between tiles travels vertically up the graphic.'' \\ \rowcolor{greyColor}
0:06 & 02 & a & ``Complexity increases vertically'' & ``The tiles at the bottom focus on vertical lines or bumpy edges while the items in the upper tiles look more like wolf faces.'' \\
0:07 & 03 & a & ``Layer 5A Mixed has better examples'' & ``The system must be using the data from this layer to get such high accuracy.'' \\ \rowcolor{greyColor}
0:08 & 01 & b & ``Data flows throughout the network'' & ``It doesn't make sense that an input image would visit each of these nodes, but rather that each node is a "filter" to apply over an input image and the larger the Node, the more important it is for determining an the class.'' \\ \hline \rowcolor{someColor}
\multicolumn{1}{|l}{0:10} & & & \multicolumn{2}{l|}{Activity: Inspecting the class ``Tench''} \\ \hdashline
0:14 & 04 & a & ``Are Tench people or fish?'' & ``I see some finger-like and human face nodes mixed in with what look like eels.'' \\ \rowcolor{greyColor}
0:15 & 04 & b & ``Tench appear to be fish'' & ``The other similar classes are things I know are fish like a barracuda or sturgeon, but this doesn't explain the finger-like and human face nodes mixed in.'' \\
0:17 & 05 & a & ``Mixed layers 4a b \& c all respond primarily to faces while 4 d and e look more like fish'' & ``Not sure why this happens but maybe the fish has a camoflauge that looks like human faces or other animals.'' \\ \rowcolor{greyColor}
0:18 & 06 & a & ``Tench have a high accuracy even though they appear to have a confusing tree'' & ``Could this mean that the system is using other features in an image of tench to help identify the fish?'' \\
0:20 & 06 & b & ``The system uses alternative features to recognize Tench'' & ``The system clearly uses features of people and fingers to recognize Tench, could it be that these images come from holding a caught fish like a prize?'' \\ \rowcolor{greyColor}
0:22 & 02 & b & ``Complexity is learned from the lower features'' & ``As a neural network, the lower layers feed the patterns recognized by the upper layers, thus more human descriptive nodes.''
\\ \hdashline
\end{tabular}}
\vspace{0.3cm}
\caption{Example from the use case illustrating data collection via the micro-entry tool. 
Each row records a micro-entry at a specific time and is given an identifying number and version letter.
Revised entries are denoted by a new letter in the Ver (version) column. 
Notice how the rationales' shift over subsequent entries, especially when an entry is revisited (Ver = b). 
This user reaches conclusions faster than to be expected in practice.}
\label{tab:entries}
\end{table*}

\section{Use Case}
To further illustrate how micro-entries reveal changes in a user's mental model, we walk through their use for collecting and interpreting data from an interactive session with an XAI application.
We present the use case based on Summit (see Figure ~\ref{fig:tool}), a visualization that summarizes a multiclass image classifier~\cite{Hohman_Park_Robinson_Chau_2019}. 
The tool provides an attribution graph to 
visualize what features a model has learned as nodes and how they are related with edges.
Additional views show a two-dimensional embedding of all classes (upper left) and a searchable list of available classes to compare (bottom left).
Summit is intended to give users an impression of the underlying neuron activations in order to find areas where the model could improve.

In the interest of facilitating communication, we've named our user Shelly, and a template design for micro-entries is offered at the right of Figure \ref{fig:tool}.
While this example describes a more open exploration task, a similar method could likely be adapted to more controlled or instance-based procedures; we leave these adaptations for future discussion. 



Shelly's task is to inspect a list of classes with the goal of helping researchers identify reasons the model may be making mistakes.
After being introduced to Summit and its functions, Shelly is asked to employ the micro-entry method while completing her task.
The researchers ask her to ``describe'' and ``justify'' each pattern she identifies while exploring a set of classes in Summit.
If she changes her mind, she's asked to update any previous entry. 
The researchers do their best to establish rapport with Shelly so she feels valued as a participant and understands how she contributes to visualization scholarship~\cite{Gomoll_Nicol_1990}.
Table \ref{tab:entries} shows a hypothetical early sub-set of her entries while working through the task.

Exploring the White Wolf class, she first notices that Summit has animated edges that turn blue as she hovers her mouse over the tree---writing about this realization in entry 01a.
As she continues, she makes a new entry (02a) for a different observed pattern: `how complexity appears to be encoded vertically.'
Information about understanding the visual representation is common when employing think-aloud and observation, but the temporal nature will allow us to see how interface learning can relate to their mental model of the computational model.

After 10 minutes of exploration, Shelly is directed to look at the \textit{tench} class.
In short order, Shelly has noticed suspicious issues involved with this class (see entry 4a) and is unsure if the class \textit{tench} should have elements of hands, faces, or fish. 
Entry 04 is updated (version b) as her mental model crystallizes along with her justification: `\textit{Tench} must be fish because \textit{barracuda} and \textit{sturgeon} are similarly classified.'
As she continues to inspect the class, she concludes that there is bias in how the system classifies the \textit{tench} class (entry 06b) because it activates neurons that are also related to human faces and hands. 

In only the first two classes, it appears that Shelly uncovers some unique patterns in the classifier using the Summit visualization.
Her entries feel focused and show signs of learning and pattern recognition instead of over-generalizations because she is asked to provide her own explanations for the patterns.
Hypothetically, if we assume additional participants completed this same task, we may expose some answers to the six questions we introduced in Section \ref{interpretations}.
For example, it would be interesting to compare how different visualizations may change what kinds of patterns are noticed and the justifications provided for them.
Alternatively, by hiding various elements in an interface, the patterns and justifications provided by micro-entries may help uncover the communicative potency of each interface element.
While the micro-entry method may not help identify specifically what is working or not working in visualizations, it could help uncover how well a visualization communicates overall.
More importantly, the method can help reveal how visual design leads to insights or changes in the user's understanding of a machine learning model.
\section{Conclusion}

We argue for increased attention to shifts in human understanding and perceptions of system capabilities in interactive data tools.
The fundamental ideas discussed in this paper pull together concepts and methods established by a rich history of existing research, but there is a clear need for deeper evaluation of mental models as research advances continue for interpretable and transparent system design~\cite{Adadi_Berrada_2018,Doshi-Velez_Kim_2017,miller2019explanation}.
There is potential to incorporate fundamental elements of reflection in all tasks, as it helps refine users' mental model and facilitate their communication.

We chose to focus on the micro-entries technique---as described in this paper---in our use case to illustrate it's potential, conceptualize how its data may look in practice, and lay a foundation for future designs.
Additionally, our methodology---similar to online journaling~\cite{Kessler_Lund_2004}---has applications to systems that require remote evaluation when study populations may be physically inaccessible (e.g., during the COVID-19 pandemic). 
Alternative methods, like think aloud or concept maps, appeal to alternative modalities that may offer unique, beneficial insights and ought to be explored~\cite{Hatami_2013}.

We conclude with three future research questions: 
\begin{itemize}
\itemsep-0.2em 

\item How do mental models shift while working with complex visualizations and micro-entries?

\item What common understandings are communicated by specific visualization features when users are asked to reflect?

\item How do micro-entries compare to other mental model measurement methods? 

\end{itemize}

From the presented use case, applying our proposed methodology to an XAI visualization and validating its effectiveness compared to alternative techniques is an appropriate exploratory domain with immediate and potentially fruitful insights.
Ultimately, as researchers in the visualization and data science communities are interested in making algorithms more interpretable, our method aims to make the data and decision boundaries more visible and accessible to the users by inviting them to reflect on what they know.
By prompting users to consider their own recognized patterns---and explain to themselves why they exist---the micro-entry method can elicit the construction of more thorough understanding while also capturing one's stages of mental model development.
These stages of development carry time-specific, user-defined interpretations; enabling generative evaluations and clearer illustrations of what a visualization communicates.
\acknowledgments{
This research was supported by NSF awards 1900767 and 1929693, and by the DARPA XAI program under N66001-17-2-4032.

}

\bibliographystyle{abbrv-doi-hyperref-narrow}

\bibliography{ms}

\begin{thebibliography}{10}

\bibitem{Adadi_Berrada_2018}
A.~Adadi and M.~Berrada.
\newblock Peeking inside the black-box: A survey on explainable artificial
  intelligence (xai).
\newblock {\em IEEE Access}, 6:52138–52160, 2018. doi: {{%
10\hspace{.1pt}\discretionary{.}{%
}{.}\hspace{.4pt}1109\discretionary{/}{%
}{/}ACCESS\hspace{.1pt}\discretionary{.}{%
}{.}\hspace{.4pt}2018\hspace{.1pt}\discretionary{.}{%
}{.}\hspace{.4pt}2870052}}


\bibitem{aigner2007visualizing}
W.~Aigner, S.~Miksch, W.~M{\"u}ller, H.~Schumann, and C.~Tominski.
\newblock Visualizing time-oriented data—a systematic view.
\newblock {\em Computers \& Graphics}, 31(3):401--409, 2007.

\bibitem{Aigner_Rind_Hoffmann_2012}
W.~Aigner, A.~Rind, and S.~Hoffmann.
\newblock Comparative evaluation of an interactive time-series visualization
  that combines quantitative data with qualitative abstractions.
\newblock {\em Computer Graphics Forum}, 31(3pt2):995--1004, 2012. doi: {{%
10\hspace{.1pt}\discretionary{.}{%
}{.}\hspace{.4pt}1111\discretionary{/}{%
}{/}j\hspace{.1pt}\discretionary{.}{%
}{.}\hspace{.4pt}1467\discretionary{%
}{-}{-}8659\hspace{.1pt}\discretionary{.}{%
}{.}\hspace{.4pt}2012\hspace{.1pt}\discretionary{.}{%
}{.}\hspace{.4pt}03092\hspace{.1pt}\discretionary{.}{%
}{.}\hspace{.4pt}x}}


\bibitem{arrieta_2020}
A.~B. Arrieta, N.~Díaz-Rodríguez, J.~D. Ser, A.~Bennetot, S.~Tabik,
  A.~Barbado, S.~Garcia, S.~Gil-Lopez, D.~Molina, R.~Benjamins, R.~chatila, and
  F.~Herrera.
\newblock Explainable artificial intelligence (xai): Concepts, taxonomies,
  opportunities and challenges toward responsible ai.
\newblock {\em Information Fusion}, 58:82–115, 2020. doi: {{%
10\hspace{.1pt}\discretionary{.}{%
}{.}\hspace{.4pt}1016\discretionary{/}{%
}{/}j\hspace{.1pt}\discretionary{.}{%
}{.}\hspace{.4pt}inffus\hspace{.1pt}\discretionary{.}{%
}{.}\hspace{.4pt}2019\hspace{.1pt}\discretionary{.}{%
}{.}\hspace{.4pt}12\hspace{.1pt}\discretionary{.}{%
}{.}\hspace{.4pt}012}}


\bibitem{Baldwin_2000}
M.~Baldwin.
\newblock Does self-assessment in a group help students to learn?
\newblock {\em Social Work Education}, 19(5):451–462, Oct 2000.

\bibitem{Bolte_Nourani_Ragan_Bruckner_2020}
F.~Bolte, M.~Nourani, E.~D. Ragan, and S.~Bruckner.
\newblock Splitstreams: A visual metaphor for evolving hierarchies.
\newblock {\em IEEE Transactions on Visualization and Computer Graphics}, p.
  1–1, 2020. doi: {{%
10\hspace{.1pt}\discretionary{.}{%
}{.}\hspace{.4pt}1109\discretionary{/}{%
}{/}TVCG\hspace{.1pt}\discretionary{.}{%
}{.}\hspace{.4pt}2020\hspace{.1pt}\discretionary{.}{%
}{.}\hspace{.4pt}2973564}}


\bibitem{Boren_Ramey_2000}
T.~Boren and J.~Ramey.
\newblock Thinking aloud: Reconciling theory and practice.
\newblock {\em Professional Communication, IEEE Transactions on}, 43:261–278,
  Oct 2000. doi: {{%
10\hspace{.1pt}\discretionary{.}{%
}{.}\hspace{.4pt}1109\discretionary{/}{%
}{/}47\hspace{.1pt}\discretionary{.}{%
}{.}\hspace{.4pt}867942}}


\bibitem{Boud_Keogh_Walker_Keogh_Walker_1985}
D.~Boud, R.~Keogh, D.~Walker, R.~Keogh, and D.~Walker.
\newblock {\em Reflection: Turning Experience into Learning}.
\newblock Routledge, 1985. doi: {{%
10\hspace{.1pt}\discretionary{.}{%
}{.}\hspace{.4pt}4324\discretionary{/}{%
}{/}9781315059051}}


\bibitem{Braun_Clarke_2006}
V.~Braun and V.~Clarke.
\newblock Using thematic analysis in psychology.
\newblock {\em Qualitative Research in Psychology}, 3(2):77–101, Jan 2006.
  doi: {{%
10\hspace{.1pt}\discretionary{.}{%
}{.}\hspace{.4pt}1191\discretionary{/}{%
}{/}1478088706qp063oa}}


\bibitem{Braun_Clarke_Hayfield_Terry_2018}
V.~Braun, V.~Clarke, N.~Hayfield, and G.~Terry.
\newblock Thematic analysis.
\newblock In P.~Liamputtong, ed., {\em Handbook of Research Methods in Health
  Social Sciences}, p. 1–18. Springer, 2018. doi: {{%
10\hspace{.1pt}\discretionary{.}{%
}{.}\hspace{.4pt}1007\discretionary{/}{%
}{/}978\discretionary{%
}{-}{-}981\discretionary{%
}{-}{-}10\discretionary{%
}{-}{-}2779\discretionary{%
}{-}{-}6\_103\discretionary{%
}{-}{-}1}}


\bibitem{Bull_Ginon_Boscolo_Johnson_2016}
S.~Bull, B.~Ginon, C.~Boscolo, and M.~Johnson.
\newblock Introduction of learning visualisations and metacognitive support in
  a persuadable open learner model.
\newblock In {\em Proceedings of the Sixth International Conference on Learning
  Analytics \& Knowledge}, LAK ’16, p. 30–39. Association for Computing
  Machinery, Apr 2016. doi: {{%
10\hspace{.1pt}\discretionary{.}{%
}{.}\hspace{.4pt}1145\discretionary{/}{%
}{/}2883851\hspace{.1pt}\discretionary{.}{%
}{.}\hspace{.4pt}2883853}}


\bibitem{Cabrera_Cabera_Sokolow_Mearis_2018}
D.~Cabrera, L.~Cabera, J.~Sokolow, and D.~Mearis.
\newblock Why you should map: the science behind visual mapping.
\newblock {\em Plectica Publishing}, 2018.

\bibitem{Cabrera_Cabrera_Troxell_2020}
D.~Cabrera, L.~Cabrera, and G.~Troxell.
\newblock The future of systems x?
\newblock {\em Journal of Applied Systems Thinking}, 20(5):1–13, May 2020.

\bibitem{Carpendale_2008}
S.~Carpendale.
\newblock {\em Evaluating Information Visualizations}, vol. 4950 of {\em
  Lecture Notes in Computer Science}, p. 19–45.
\newblock Springer Berlin Heidelberg, 2008. doi: {{%
10\hspace{.1pt}\discretionary{.}{%
}{.}\hspace{.4pt}1007\discretionary{/}{%
}{/}978\discretionary{%
}{-}{-}3\discretionary{%
}{-}{-}540\discretionary{%
}{-}{-}70956\discretionary{%
}{-}{-}5\_2}}


\bibitem{Carroll_Olson_1987}
J.~M. Carroll and J.~R. Olson.
\newblock {\em Mental Models in Human-Computer Interaction. Research Issues
  about What the User of Software Knows. Workshop on Software Human Factors:
  Users’ Mental Models (Washington, District of Columbia, May 15-16, 1984)}.
\newblock Committee on Human Factors, Commission on Behavioral and Social
  Sciences and Education, National Research Council, 2101 Constitution Ave,
  1987.

\bibitem{Cheshire_Muldoon_Francis_Lewis_Ball_2007}
A.~Cheshire, K.~P. Muldoon, B.~Francis, C.~N. Lewis, and L.~J. Ball.
\newblock Modelling change: new opportunities in the analysis of microgenetic
  data.
\newblock {\em Infant and Child Development}, 16(1):119–134, Feb 2007. doi:
  {{%
10\hspace{.1pt}\discretionary{.}{%
}{.}\hspace{.4pt}1002\discretionary{/}{%
}{/}icd\hspace{.1pt}\discretionary{.}{%
}{.}\hspace{.4pt}498}}


\bibitem{Chi_2013}
M.~T.~H. Chi.
\newblock {\em Two Kinds and Four Sub-types of Misconceived Knowledge, Ways to
  Change it, and the Learning Outcomes}, vol.~2, p.~22.
\newblock Routledge, 2013.

\bibitem{Craik_1943}
K.~J.~W. Craik.
\newblock {\em The Nature of Explanation}.
\newblock Cambridge University Press, Jan 1943.
\newblock Google-Books-ID: EN0TrgEACAAJ.

\bibitem{cummings2004automation}
M.~Cummings.
\newblock Automation bias in intelligent time critical decision support
  systems.
\newblock In {\em AIAA 1st Intelligent Systems Technical Conference}, p. 6313,
  2004.

\bibitem{dasgupta2016familiarity}
A.~Dasgupta, J.-Y. Lee, R.~Wilson, R.~A. Lafrance, N.~Cramer, K.~Cook, and
  S.~Payne.
\newblock Familiarity vs trust: A comparative study of domain scientists' trust
  in visual analytics and conventional analysis methods.
\newblock {\em IEEE transactions on visualization and computer graphics},
  23(1):271--280, 2016.

\bibitem{Denning_Hoiem_Simpson_Sullivan_1990}
S.~Denning, D.~Hoiem, M.~Simpson, and K.~Sullivan.
\newblock The value of thinking-aloud protocols in industry: A case study at
  microsoft corporation.
\newblock {\em Proceedings of the Human Factors Society Annual Meeting},
  34(17):1285–1289, Oct 1990. doi: {{%
10\hspace{.1pt}\discretionary{.}{%
}{.}\hspace{.4pt}1177\discretionary{/}{%
}{/}154193129003401723}}


\bibitem{dietvorst2016people}
B.~Dietvorst.
\newblock People reject (superior) algorithms because they compare them to
  counter-normative reference points.
\newblock {\em Available at SSRN 2881503}, 2016.

\bibitem{Doshi-Velez_Kim_2017}
F.~Doshi-Velez and B.~Kim.
\newblock Towards a rigorous science of interpretable machine learning.
\newblock {\em arXiv:1702.08608 [cs, stat]}, Mar 2017.
\newblock arXiv: 1702.08608.

\bibitem{Edwards_2000}
D.~Edwards.
\newblock {\em Introduction to graphical modelling}.
\newblock Springer, 2000.

\bibitem{Ellis_Dix_2006}
G.~Ellis and A.~Dix.
\newblock An explorative analysis of user evaluation studies in information
  visualisation.
\newblock In {\em Proceedings of the 2006 AVI Workshop on BEyond Time and
  Errors: Novel Evaluation Methods for Information Visualization}, BELIV '06,
  p. 1–7. Association for Computing Machinery, New York, NY, USA, 2006. doi:
  {{%
10\hspace{.1pt}\discretionary{.}{%
}{.}\hspace{.4pt}1145\discretionary{/}{%
}{/}1168149\hspace{.1pt}\discretionary{.}{%
}{.}\hspace{.4pt}1168152}}


\bibitem{Ericsson_Simon_1984}
K.~A. Ericsson and H.~A. Simon.
\newblock {\em Protocol analysis: Verbal reports as data}.
\newblock Protocol analysis: Verbal reports as data. The MIT Press, 1984.

\bibitem{Evans_Jentsch_Hitt_Bowers_Salas_2001}
A.~W. Evans, F.~Jentsch, J.~M. Hitt, C.~Bowers, and E.~Salas.
\newblock Mental model assessments: Is there convergence among different
  methods?
\newblock {\em Proceedings of the Human Factors and Ergonomics Society Annual
  Meeting}, 45(4):293–296, Oct 2001. doi: {{%
10\hspace{.1pt}\discretionary{.}{%
}{.}\hspace{.4pt}1177\discretionary{/}{%
}{/}154193120104500406}}


\bibitem{Flynn_Pine_Lewis_2006}
E.~Flynn, K.~Pine, and C.~Lewis.
\newblock The microgenetic method - "time for change?".
\newblock {\em The Psychologist}, 19(3):4, Mar 2006.

\bibitem{Foraita_Klasen_Pigeot_2008}
R.~Foraita, S.~Klasen, and I.~Pigeot.
\newblock Using graphical chain models to analyze differences in structural
  correlates of undernutrition in benin and bangladesh.
\newblock {\em Economics \& Human Biology}, 6(3):398–419, Dec 2008. doi: {{%
10\hspace{.1pt}\discretionary{.}{%
}{.}\hspace{.4pt}1016\discretionary{/}{%
}{/}j\hspace{.1pt}\discretionary{.}{%
}{.}\hspace{.4pt}ehb\hspace{.1pt}\discretionary{.}{%
}{.}\hspace{.4pt}2008\hspace{.1pt}\discretionary{.}{%
}{.}\hspace{.4pt}07\hspace{.1pt}\discretionary{.}{%
}{.}\hspace{.4pt}002}}


\bibitem{Gan_Zhu_Li_Liang_Cao_Zhou_2014}
Q.~Gan, M.~Zhu, M.~Li, T.~Liang, Y.~Cao, and B.~Zhou.
\newblock Document visualization: an overview of current research.
\newblock {\em WIREs Computational Statistics}, 6(1):19–36, 2014. doi: {{%
10\hspace{.1pt}\discretionary{.}{%
}{.}\hspace{.4pt}1002\discretionary{/}{%
}{/}wics\hspace{.1pt}\discretionary{.}{%
}{.}\hspace{.4pt}1285}}


\bibitem{Garcia_Chu_Nam_Banigan_2018}
B.~Garcia, S.~L. Chu, B.~Nam, and C.~Banigan.
\newblock Wearables for learning: Examining the smartwatch as a tool for
  situated science reflection.
\newblock In {\em Proceedings of the 2018 CHI Conference on Human Factors in
  Computing Systems - CHI ’18}, p. 1–13. ACM Press, 2018. doi: {{%
10\hspace{.1pt}\discretionary{.}{%
}{.}\hspace{.4pt}1145\discretionary{/}{%
}{/}3173574\hspace{.1pt}\discretionary{.}{%
}{.}\hspace{.4pt}3173830}}


\bibitem{GlazerWaldman_Cox_1980}
H.~Glazer-Waldman and D.~L. Cox.
\newblock The use of similarity judgments to assess the effectiveness of
  instruction.
\newblock {\em Education}, 100(4):352–59, 1980.

\bibitem{Gomoll_Nicol_1990}
K.~Gomoll and A.~Nicol.
\newblock User observation: Guidelines for apple developers, Jan 1990.

\bibitem{goodall2018situ}
J.~R. Goodall, E.~D. Ragan, C.~A. Steed, J.~W. Reed, G.~D. Richardson, K.~M.
  Huffer, R.~A. Bridges, and J.~A. Laska.
\newblock Situ: Identifying and explaining suspicious behavior in networks.
\newblock {\em IEEE transactions on visualization and computer graphics},
  25(1):204--214, 2018.

\bibitem{Hadinh_Bonner_Clark_Ramsbotham_Hines_2016}
T.~T. Ha~Dinh, A.~Bonner, R.~Clark, J.~Ramsbotham, and S.~Hines.
\newblock The effectiveness of the teach-back method on adherence and
  self-management in health education for people with chronic disease: a
  systematic review.
\newblock {\em JBI Evidence Synthesis}, 14(1):210–247, Jan 2016. doi: {{%
10\hspace{.1pt}\discretionary{.}{%
}{.}\hspace{.4pt}11124\discretionary{/}{%
}{/}jbisrir\discretionary{%
}{-}{-}2016\discretionary{%
}{-}{-}2296}}


\bibitem{Hampton_Morrow_2003}
S.~E. Hampton and C.~Morrow.
\newblock Reflective journaling and assessment.
\newblock {\em Journal of Professional Issues in Engineering Education and
  Practice}, 129(4):186–189, Oct 2003. doi: {{%
10\hspace{.1pt}\discretionary{.}{%
}{.}\hspace{.4pt}1061\discretionary{/}{%
}{/}\discretionary{%
}{(}{(}ASCE\discretionary{)}{%
}{)}1052\discretionary{%
}{-}{-}3928\discretionary{%
}{(}{(}2003\discretionary{)}{%
}{)}129\discretionary{:}{%
}{:}4\discretionary{%
}{(}{(}186)}}


\bibitem{Hardiman_Dufresne_Mestre_1989}
P.~T. Hardiman, R.~Dufresne, and J.~P. Mestre.
\newblock The relation between problem categorization and problem solving among
  experts and novices.
\newblock {\em Memory \& Cognition}, 17(5):627–638, Sep 1989. doi: {{%
10\hspace{.1pt}\discretionary{.}{%
}{.}\hspace{.4pt}3758\discretionary{/}{%
}{/}BF03197085}}


\bibitem{Hart-Davidson_Spinuzzi_Zachry_2007}
W.~Hart-Davidson, C.~Spinuzzi, and M.~Zachry.
\newblock Capturing \& visualizing knowledge work: Results \& implications of a
  pilot study of proposal writing activity.
\newblock {\em Association for Computing Machinery}, p. 113–119, Oct 2007.
  doi: {{%
10\hspace{.1pt}\discretionary{.}{%
}{.}\hspace{.4pt}1145\discretionary{/}{%
}{/}1297144\hspace{.1pt}\discretionary{.}{%
}{.}\hspace{.4pt}1297168}}


\bibitem{Hatami_2013}
S.~Hatami.
\newblock Learning styles.
\newblock {\em ELT Journal}, 67(4):488–490, Oct 2013. doi: {{%
10\hspace{.1pt}\discretionary{.}{%
}{.}\hspace{.4pt}1093\discretionary{/}{%
}{/}elt\discretionary{/}{%
}{/}ccs083}}


\bibitem{Havre_Hetzler_Nowell_2000}
S.~Havre, B.~Hetzler, and L.~Nowell.
\newblock Themeriver: visualizing theme changes over time.
\newblock In {\em IEEE Symposium on Information Visualization 2000. INFOVIS
  2000. Proceedings}, p. 115–123. IEEE Comput. Soc, 2000. doi: {{%
10\hspace{.1pt}\discretionary{.}{%
}{.}\hspace{.4pt}1109\discretionary{/}{%
}{/}INFVIS\hspace{.1pt}\discretionary{.}{%
}{.}\hspace{.4pt}2000\hspace{.1pt}\discretionary{.}{%
}{.}\hspace{.4pt}885098}}


\bibitem{Heimerl_Lohmann_Lange_Ertl_2014}
F.~Heimerl, S.~Lohmann, S.~Lange, and T.~Ertl.
\newblock Word cloud explorer: Text analytics based on word clouds.
\newblock In {\em 2014 47th Hawaii International Conference on System
  Sciences}, p. 1833–1842, Jan 2014. doi: {{%
10\hspace{.1pt}\discretionary{.}{%
}{.}\hspace{.4pt}1109\discretionary{/}{%
}{/}HICSS\hspace{.1pt}\discretionary{.}{%
}{.}\hspace{.4pt}2014\hspace{.1pt}\discretionary{.}{%
}{.}\hspace{.4pt}231}}


\bibitem{Herman_2019}
B.~Herman.
\newblock The promise and peril of human evaluation for model interpretability.
\newblock {\em arXiv:1711.07414 [cs, stat]}, Oct 2019.
\newblock arXiv: 1711.07414.

\bibitem{herschel2017survey}
M.~Herschel, R.~Diestelk{\"a}mper, and H.~B. Lahmar.
\newblock A survey on provenance: What for? what form? what from?
\newblock {\em The VLDB Journal}, 26(6):881--906, 2017.

\bibitem{Hoffman_Mueller_Klein_Litman_2019}
R.~R. Hoffman, S.~T. Mueller, G.~Klein, and J.~Litman.
\newblock Metrics for explainable ai: Challenges and prospects.
\newblock {\em arXiv:1812.04608 [cs]}, Feb 2019.
\newblock arXiv: 1812.04608.

\bibitem{Hohman_Park_Robinson_Chau_2019}
F.~Hohman, H.~Park, C.~Robinson, and D.~H. Chau.
\newblock Summit: Scaling deep learning interpretability by visualizing
  activation and attribution summarizations.
\newblock {\em IEEE Transactions on Visualization and Computer Graphics}, p.
  1–1, Aug 2019. doi: {{%
10\hspace{.1pt}\discretionary{.}{%
}{.}\hspace{.4pt}1109\discretionary{/}{%
}{/}TVCG\hspace{.1pt}\discretionary{.}{%
}{.}\hspace{.4pt}2019\hspace{.1pt}\discretionary{.}{%
}{.}\hspace{.4pt}2934659}}


\bibitem{honeycutt2020solicit}
D.~R. Honeycutt, M.~Nourani, and E.~D. Ragan.
\newblock Soliciting human-in-the-loop user feedback for interactive machine
  learning reduces user trust and impressions of model accuracy.
\newblock In {\em Eighth AAAI Conference on Human Computation and
  Crowdsourcing}, 2020.

\bibitem{Hosanagar_Jair_2018}
K.~Hosanagar and V.~Jair.
\newblock We need transparency in algorithms, but too much can backfire.
\newblock {\em Harvard Business Review}, Jul 2018.

\bibitem{Hung_Blumschein_2009}
W.~Hung and P.~Blumschein.
\newblock {\em After word: Where do we go from here?}, p. 319–329.
\newblock Sense, Jan 2009. doi: {{%
10\hspace{.1pt}\discretionary{.}{%
}{.}\hspace{.4pt}1163\discretionary{/}{%
}{/}9789087907112\_020}}


\bibitem{Isaacson_Fujita_2006}
R.~M. Isaacson and F.~Fujita.
\newblock Metacognitive knowledge monitoring and self-regulated learning:
  Academic success and reflections on learning.
\newblock {\em Journal of the Scholarship of Teaching and Learning},
  6(1):39–55, Aug 2006.

\bibitem{Johnson-Laird_2010}
P.~N. Johnson-Laird.
\newblock Mental models and human reasoning.
\newblock {\em Proceedings of the National Academy of Sciences}, 107(43):8, Oct
  2010. doi: {{%
10\hspace{.1pt}\discretionary{.}{%
}{.}\hspace{.4pt}1073\discretionary{/}{%
}{/}pnas\hspace{.1pt}\discretionary{.}{%
}{.}\hspace{.4pt}1012933107}}


\bibitem{Kessler_Lund_2004}
P.~D. Kessler and C.~H. Lund.
\newblock Reflective journaling: Developing an online journal for distance
  education.
\newblock {\em Nurse Educator}, 29(1):20–24, Feb 2004.

\bibitem{Khemlani_Johnson-Laird_2011}
S.~S. Khemlani and P.~N. Johnson-Laird.
\newblock The need to explain.
\newblock {\em Quarterly Journal of Experimental Psychology},
  64(11):2276–2288, Nov 2011. doi: {{%
10\hspace{.1pt}\discretionary{.}{%
}{.}\hspace{.4pt}1080\discretionary{/}{%
}{/}17470218\hspace{.1pt}\discretionary{.}{%
}{.}\hspace{.4pt}2011\hspace{.1pt}\discretionary{.}{%
}{.}\hspace{.4pt}592593}}


\bibitem{Klein_Militello_2001}
G.~Klein and L.~Militello.
\newblock {\em Some guidelines for conducting a cognitive task analysis},
  vol.~1, p. 163–199.
\newblock Emerald (MCB UP ), 2001. doi: {{%
10\hspace{.1pt}\discretionary{.}{%
}{.}\hspace{.4pt}1016\discretionary{/}{%
}{/}S1479\discretionary{%
}{-}{-}3601\discretionary{%
}{(}{(}01\discretionary{)}{%
}{)}01006\discretionary{%
}{-}{-}2}}


\bibitem{Kulesza_Stumpf_Burnett_Wong_Riche_Moore_Oberst_Shinsel_McIntosh_2010}
T.~Kulesza, S.~Stumpf, M.~Burnett, W.-K. Wong, Y.~Riche, T.~Moore, I.~Oberst,
  A.~Shinsel, and K.~McIntosh.
\newblock Explanatory debugging: Supporting end-user debugging of
  machine-learned programs.
\newblock In {\em 2010 IEEE Symposium on Visual Languages and Human-Centric
  Computing}, p. 41–48. IEEE, Sep 2010. doi: {{%
10\hspace{.1pt}\discretionary{.}{%
}{.}\hspace{.4pt}1109\discretionary{/}{%
}{/}VLHCC\hspace{.1pt}\discretionary{.}{%
}{.}\hspace{.4pt}2010\hspace{.1pt}\discretionary{.}{%
}{.}\hspace{.4pt}15}}


\bibitem{Kuusela_Paul_2000}
H.~Kuusela and P.~Paul.
\newblock A comparison of concurrent and retrospective verbal protocol
  analysis.
\newblock {\em The American Journal of Psychology}, 113(3):387–404, 2000.
  doi: {{%
10\hspace{.1pt}\discretionary{.}{%
}{.}\hspace{.4pt}2307\discretionary{/}{%
}{/}1423365}}


\bibitem{Lam_Bertini_Isenberg_Plaisant_Carpendale_2011}
H.~Lam, E.~Bertini, P.~Isenberg, C.~Plaisant, and S.~Carpendale.
\newblock Seven guiding scenarios for information visualization evaluation,
  2011.

\bibitem{Lipton_2017}
Z.~C. Lipton.
\newblock The mythos of model interpretability.
\newblock {\em arXiv:1606.03490 [cs, stat]}, Mar 2017.
\newblock arXiv: 1606.03490.

\bibitem{Luwel_2012}
K.~Luwel.
\newblock Microgenetic method.
\newblock In N.~M. Seel, ed., {\em Encyclopedia of the Sciences of Learning},
  p. 2265–2268. Springer US, 2012. doi: {{%
10\hspace{.1pt}\discretionary{.}{%
}{.}\hspace{.4pt}1007\discretionary{/}{%
}{/}978\discretionary{%
}{-}{-}1\discretionary{%
}{-}{-}4419\discretionary{%
}{-}{-}1428\discretionary{%
}{-}{-}6\_1754}}


\bibitem{Merton_Kendall_1946}
R.~K. Merton and P.~L. Kendall.
\newblock The focused interview.
\newblock {\em American Journal of Sociology}, 51(6):541–557, May 1946. doi:
  {{%
10\hspace{.1pt}\discretionary{.}{%
}{.}\hspace{.4pt}1086\discretionary{/}{%
}{/}219886}}


\bibitem{miller2019explanation}
T.~Miller.
\newblock Explanation in artificial intelligence: Insights from the social
  sciences.
\newblock {\em Artificial Intelligence}, 267:1--38, 2019.

\bibitem{miller2017explainable}
T.~Miller, P.~Howe, and L.~Sonenberg.
\newblock Explainable ai: Beware of inmates running the asylum or: How i learnt
  to stop worrying and love the social and behavioural sciences.
\newblock {\em arXiv preprint arXiv:1712.00547}, 2017.

\bibitem{Milrad_Spector_Davidsen_2002}
M.~Milrad, J.~M. Spector, P.~I. Davidsen, et~al.
\newblock Model facilitated learning.
\newblock {\em Learning and teaching with technology: Principles and
  practices}, pp. 13--27, 2003.

\bibitem{Mohseni_Zarei_Ragan_2018}
S.~Mohseni, N.~Zarei, and E.~D. Ragan.
\newblock A survey of evaluation methods and measures for interpretable machine
  learning.
\newblock {\em arXiv:1811.11839 [cs]}, Dec 2018.
\newblock arXiv: 1811.11839.

\bibitem{Nakatsu_2009_3}
R.~T. Nakatsu.
\newblock {\em Chapter 3: TYPES OF DIAGRAMS}, p. 346.
\newblock John Wiley \& Sons, Incorpated, 1 ed., Dec 2009.

\bibitem{Norman_1983}
D.~A. Norman.
\newblock {\em Some Observations on Mental Models}, p. 7–14.
\newblock Lawrence Erlbaum Associates Inc. pp7-14, 1 ed., 1983.

\bibitem{North_Chang_Endert_Dou_2011}
C.~North, R.~Chang, A.~Endert, W.~Dou, R.~May, B.~Pike, and G.~Fink.
\newblock Analytic provenance: Process+interaction+insight.
\newblock In {\em CHI '11 Extended Abstracts on Human Factors in Computing
  Systems}, CHI EA '11, p. 33–36. Association for Computing Machinery, New
  York, NY, USA, 2011. doi: {{%
10\hspace{.1pt}\discretionary{.}{%
}{.}\hspace{.4pt}1145\discretionary{/}{%
}{/}1979742\hspace{.1pt}\discretionary{.}{%
}{.}\hspace{.4pt}1979570}}


\bibitem{Nourani_Honeycutt_Block_Roy_Rahman_Ragan_Gogate_2020}
M.~Nourani, D.~R. Honeycutt, J.~E. Block, C.~Roy, T.~Rahman, E.~D. Ragan, and
  V.~Gogate.
\newblock Investigating the importance of first impressions and explainable ai
  with interactive video analysis.
\newblock {\em ACM CHI}, p.~8, 2020.

\bibitem{nourani2019effects}
M.~Nourani, S.~Kabir, S.~Mohseni, and E.~D. Ragan.
\newblock The effects of meaningful and meaningless explanations on trust and
  perceived system accuracy in intelligent systems.
\newblock In {\em Proceedings of the AAAI Conference on Human Computation and
  Crowdsourcing}, vol.~7, pp. 97--105, 2019.

\bibitem{nourani2020Role}
M.~Nourani, J.~T. King, and E.~D. Ragan.
\newblock The role of domain expertise in user trust and the impact of first
  impressions with intelligent systems.
\newblock In {\em Eighth AAAI Conference on Human Computation and
  Crowdsourcing}, 2020.

\bibitem{parasuraman1997humans}
R.~Parasuraman and V.~Riley.
\newblock Humans and automation: Use, misuse, disuse, abuse.
\newblock {\em Human factors}, 39(2):230--253, 1997.

\bibitem{Pirolli_Card_2005}
P.~Pirolli and S.~Card.
\newblock The sensemaking process and leverage points for analyst technology as
  identified through cognitive task analysis.
\newblock In {\em International Conference on Intelligence Analysis}, Jan 2005.

\bibitem{Poursabzi-Sangdeh_Goldstein_Hofman_Vaughan_Wallach_2019}
F.~Poursabzi-Sangdeh, D.~G. Goldstein, J.~M. Hofman, J.~W. Vaughan, and
  H.~Wallach.
\newblock Manipulating and measuring model interpretability.
\newblock {\em arXiv:1802.07810 [cs]}, Nov 2019.
\newblock arXiv: 1802.07810.

\bibitem{Puerta-Melguizo_Chisalita_VanderVeer_2002}
M.~Puerta-Melguizo, C.~Chisalita, and G.~Van~der Veer.
\newblock Assessing users mental models in designing complex systems.
\newblock In {\em IEEE International Conference on Systems, Man and
  Cybernetics}, vol.~7, p.~6, Oct 2002. doi: {{%
10\hspace{.1pt}\discretionary{.}{%
}{.}\hspace{.4pt}1109\discretionary{/}{%
}{/}ICSMC\hspace{.1pt}\discretionary{.}{%
}{.}\hspace{.4pt}2002\hspace{.1pt}\discretionary{.}{%
}{.}\hspace{.4pt}1175734}}


\bibitem{ragan2015characterizing}
E.~D. Ragan, A.~Endert, J.~Sanyal, and J.~Chen.
\newblock Characterizing provenance in visualization and data analysis: an
  organizational framework of provenance types and purposes.
\newblock {\em IEEE transactions on visualization and computer graphics},
  22(1):31--40, 2015.

\bibitem{Ragan_Goodall_2014}
E.~D. Ragan and J.~R. Goodall.
\newblock Evaluation methodology for comparing memory and communication of
  analytic processes in visual analytics.
\newblock In {\em Proceedings of the Fifth Workshop on Beyond Time and Errors
  Novel Evaluation Methods for Visualization - BELIV ’14}, p. 27–34. ACM
  Press, 2014. doi: {{%
10\hspace{.1pt}\discretionary{.}{%
}{.}\hspace{.4pt}1145\discretionary{/}{%
}{/}2669557\hspace{.1pt}\discretionary{.}{%
}{.}\hspace{.4pt}2669563}}


\bibitem{ragan2015evaluating}
E.~D. Ragan, J.~R. Goodall, and A.~Tung.
\newblock Evaluating how level of detail of visual history affects process
  memory.
\newblock In {\em Proceedings of the 33rd Annual ACM Conference on Human
  Factors in Computing Systems}, pp. 2711--2720, 2015.

\bibitem{Reda_Johnson_Papka_Leigh_2016}
K.~Reda, A.~E. Johnson, M.~E. Papka, and J.~Leigh.
\newblock Modeling and evaluating user behavior in exploratory visual analysis.
\newblock {\em Information Visualization}, 15(4):325–339, Oct 2016. doi: {{%
10\hspace{.1pt}\discretionary{.}{%
}{.}\hspace{.4pt}1177\discretionary{/}{%
}{/}1473871616638546}}


\bibitem{Rouse_Morris_1985}
W.~B. Rouse and N.~M. Morris.
\newblock On looking into the black box: Prospects and limits in the search for
  mental models.
\newblock {\em Psychology Bulletin}, 100(3):349–363, May 1985.

\bibitem{Rozenblit_Keil_2002}
L.~Rozenblit and F.~Keil.
\newblock The misunderstood limits of folk science: an illusion of explanatory
  depth.
\newblock {\em Cognitive science}, 26(5):521–562, Sep 2002. doi: {{%
10\hspace{.1pt}\discretionary{.}{%
}{.}\hspace{.4pt}1207\discretionary{/}{%
}{/}s15516709cog2605\_1}}


\bibitem{Russo_Johnson_Stephens_1989}
J.~E. Russo, E.~J. Johnson, and D.~L. Stephens.
\newblock The validity of verbal protocols.
\newblock {\em Memory \& Cognition}, 17(6):759–769, Nov 1989. doi: {{%
10\hspace{.1pt}\discretionary{.}{%
}{.}\hspace{.4pt}3758\discretionary{/}{%
}{/}BF03202637}}


\bibitem{Saraiya_North_VyLam_Duca_2006}
P.~Saraiya, C.~North, V.~Lam, and K.~Duca.
\newblock An insight-based longitudinal study of visual analytics.
\newblock {\em IEEE Transactions on Visualization and Computer Graphics},
  12(6):1511–1522, Nov 2006. doi: {{%
10\hspace{.1pt}\discretionary{.}{%
}{.}\hspace{.4pt}1109\discretionary{/}{%
}{/}TVCG\hspace{.1pt}\discretionary{.}{%
}{.}\hspace{.4pt}2006\hspace{.1pt}\discretionary{.}{%
}{.}\hspace{.4pt}85}}


\bibitem{Schaffer_ODonovan_Michaelis_Raglin_Hollerer_2019}
J.~Schaffer, J.~O’Donovan, J.~Michaelis, A.~Raglin, and T.~Höllerer.
\newblock I can do better than your ai: expertise and explanations.
\newblock In {\em Proceedings of the 24th International Conference on
  Intelligent User Interfaces - IUI ’19}, p. 240–251. ACM Press, 2019. doi:
  {{%
10\hspace{.1pt}\discretionary{.}{%
}{.}\hspace{.4pt}1145\discretionary{/}{%
}{/}3301275\hspace{.1pt}\discretionary{.}{%
}{.}\hspace{.4pt}3302308}}


\bibitem{Schmidt_Biessmann_2019}
P.~Schmidt and F.~Biessmann.
\newblock Quantifying interpretability and trust in machine learning systems.
\newblock {\em arXiv:1901.08558 [cs, stat]}, Jan 2019.

\bibitem{Siegler_2002}
R.~S. Siegler.
\newblock Microgenetic studies of self-explanation.
\newblock {\em Microdevelopment: Transition Process in Development and
  Learning}, p. 31–58, May 2002. doi: {{%
10\hspace{.1pt}\discretionary{.}{%
}{.}\hspace{.4pt}1017\discretionary{/}{%
}{/}CBO9780511489709\hspace{.1pt}\discretionary{.}{%
}{.}\hspace{.4pt}002}}


\bibitem{Siegler_Crowley_1991}
R.~S. Siegler and K.~Crowley.
\newblock The microgenetic method: A direct means for studying cognitive
  development.
\newblock {\em American Psychologist}, 46(6):606–620, Jun 1991. doi: {{%
10\hspace{.1pt}\discretionary{.}{%
}{.}\hspace{.4pt}1037\discretionary{/}{%
}{/}0003\discretionary{%
}{-}{-}066X\hspace{.1pt}\discretionary{.}{%
}{.}\hspace{.4pt}46\hspace{.1pt}\discretionary{.}{%
}{.}\hspace{.4pt}6\hspace{.1pt}\discretionary{.}{%
}{.}\hspace{.4pt}606}}


\bibitem{Slone_2009}
D.~J. Slone.
\newblock Visualizing qualitative information.
\newblock {\em The Qualitative Report; Fort Lauderdale}, 14(3):489–497, Sep
  2009.

\bibitem{Smith_Corrigan_2018}
P.~Smith and G.~Corrigan.
\newblock How learners learn: A new microanalytic assessment method to map
  decision-making.
\newblock {\em Medical Teacher}, 40(12):1231–1239, Dec 2018. doi: {{%
10\hspace{.1pt}\discretionary{.}{%
}{.}\hspace{.4pt}1080\discretionary{/}{%
}{/}0142159X\hspace{.1pt}\discretionary{.}{%
}{.}\hspace{.4pt}2018\hspace{.1pt}\discretionary{.}{%
}{.}\hspace{.4pt}1426838}}


\bibitem{vanderVeer_Puerta_Melguizo_2002}
G.~C. van~der Veer and M.~d.~C. Puerta~Melguizo.
\newblock {\em Mental Models}, p. 52–60.
\newblock CRC Press, Mar 2002.

\bibitem{Viegas_Wattenberg_Dave_2004}
F.~B. Viégas, M.~Wattenberg, and K.~Dave.
\newblock Studying cooperation and conflict between authors with history flow
  visualizations.
\newblock In {\em Proceedings of the 2004 conference on Human factors in
  computing systems - CHI ’04}, p. 575–582. ACM Press, 2004. doi: {{%
10\hspace{.1pt}\discretionary{.}{%
}{.}\hspace{.4pt}1145\discretionary{/}{%
}{/}985692\hspace{.1pt}\discretionary{.}{%
}{.}\hspace{.4pt}985765}}


\bibitem{Vrotsou_Ellegard_Cooper_2007}
K.~Vrotsou, K.~Ellegard, and M.~Cooper.
\newblock Everyday life discoveries: Mining and visualizing activity patterns
  in social science diary data.
\newblock In {\em 2007 11th International Conference Information Visualization
  (IV ’07)}, p. 130–138, Jul 2007. doi: {{%
10\hspace{.1pt}\discretionary{.}{%
}{.}\hspace{.4pt}1109\discretionary{/}{%
}{/}IV\hspace{.1pt}\discretionary{.}{%
}{.}\hspace{.4pt}2007\hspace{.1pt}\discretionary{.}{%
}{.}\hspace{.4pt}48}}


\bibitem{Walker_2006}
S.~E. Walker.
\newblock Journal writing as a teaching technique to promote reflection.
\newblock {\em Journal of Athletic Training}, 41(2):216–221, 2006.

\bibitem{Wang_Xiao_Liu_Xu_Zhou_Zhang_2013}
C.~Wang, Z.~Xiao, Y.~Liu, Y.~Xu, A.~Zhou, and K.~Zhang.
\newblock Sentiview: Sentiment analysis and visualization for internet popular
  topics.
\newblock {\em IEEE Transactions on Human-Machine Systems}, 43(6):620–630,
  Nov 2013. doi: {{%
10\hspace{.1pt}\discretionary{.}{%
}{.}\hspace{.4pt}1109\discretionary{/}{%
}{/}THMS\hspace{.1pt}\discretionary{.}{%
}{.}\hspace{.4pt}2013\hspace{.1pt}\discretionary{.}{%
}{.}\hspace{.4pt}2285047}}


\bibitem{Wasserman_Koban_2019}
J.~Wasserman and K.~Koban.
\newblock Bugs on the brain: A mental model matching approach to cognitive
  skill acquisition in a strategy game.
\newblock {\em Journal of Expertise}, 2:121–139, Jun 2019.

\end{thebibliography}
\end{document}